\documentclass[10pt,a4paper,twocolumn,english,twocolumns,prl]{revtex4}
\usepackage[T1]{fontenc}
\usepackage[latin9]{inputenc}
\usepackage{amsmath}
\usepackage{graphicx}

\makeatletter

\pdfpageheight\paperheight
\pdfpagewidth\paperwidth

\@ifundefined{textcolor}{}
{%
 \definecolor{BLACK}{gray}{0}
 \definecolor{WHITE}{gray}{1}
 \definecolor{RED}{rgb}{1,0,0}
 \definecolor{GREEN}{rgb}{0,1,0}
 \definecolor{BLUE}{rgb}{0,0,1}
 \definecolor{CYAN}{cmyk}{1,0,0,0}
 \definecolor{MAGENTA}{cmyk}{0,1,0,0}
 \definecolor{YELLOW}{cmyk}{0,0,1,0}
 }

\makeatother

\usepackage{babel}
\begin{document}

\title{Modulated Floquet Topological Insulators}

\author{Yaniv Tenenbaum Katan and Daniel Podolsky}

\affiliation{Physics Department, Technion -- Israel Institute of Technology, Haifa
32000, Israel}
\begin{abstract}
The application of spatially-uniform light on conventional insulators
can induce Floquet spectra with characteristics akin to those of topological
insulators. We demonstrate that spatial modulation of light allows
for remarkable control of the properties in these systems. We provide
configurations to generate one-dimensional bulk modes, photoinduced
currents, as well as fractionalized excitations. We show a close analogy
to p-wave superconductors and use this analogy to explain our results. 
\end{abstract}
\maketitle
The Quantum Hall Effect \cite{QHE} lead to the discovery of a close
connection between topology and certain physical properties of condensed
matter systems \cite{Avron,Thouless,Haldane}. Our understanding of
the role of topology has greatly expanded following the recent discovery
of new classes of topological phases and of new materials displaying
topological properties \cite{KaneMele,Qi Zhang,Fu_Kane,Bernevig,Konig Wiedrmann,Xia Qian Hsieh,Hsieh Qian,Hassan_Kane}.
Topological phases are characterized by integer-valued numbers that
are invariant to small changes of their Hamiltonian. This makes intriguing
effects, such as quantized Hall conductivity and non-abelian excitations,
robust properties of these systems \cite{Nayak,Moore_Read,Moore,QHE}.

Recently, it has been shown that topological properties can be induced
in conventional insulators by the application of time-periodic perturbations
\cite{Kitagawa_Berg_Demler_Rudner_033429,Kitagawa_Berg_Demler_Rudner_235114,Kitagawa_Oka_Brataas_Fu_Demler,Fertig Gu Arovas Auerbach,LRG,Lindner3d,Jiang}.
Proposals for these so-called {}``Floquet Topological Insulators''
(FTIs) include a wide range of physical solid state and atomic realizations,
driven both at resonance and off-resonance. These systems display
metallic conduction enabled by quasi-stationary states at the edges
\cite{Fertig Gu Arovas Auerbach,Kitagawa_Oka_Brataas_Fu_Demler,LRG},
Dirac cones in three dimensional systems \cite{Lindner3d}, and even
Floquet Majorana fermions \cite{Jiang}.

In this Letter, we demonstrate dramatic effects that arise in FTIs
when light is modulated in space. Non-uniform light can give rise
to controlled one-dimensional modes in the bulk, to fractionalized
excitations, and to photoinduced electric currents. We establish these
results both numerically and analytically. We show that the Floquet
spectrum resembles that of a $p$-wave superconductor with a spatially-modulated
order parameter. This analogy provides a simple description of the
mechanism behind our results. We propose setups by which the properties
of light-induced topological phases can be controlled. For example,
by modifying the angle of incident light on a system one can set the
density of one-dimensional modes in its bulk.

\begin{figure}[b]
\includegraphics[width=0.4\columnwidth]{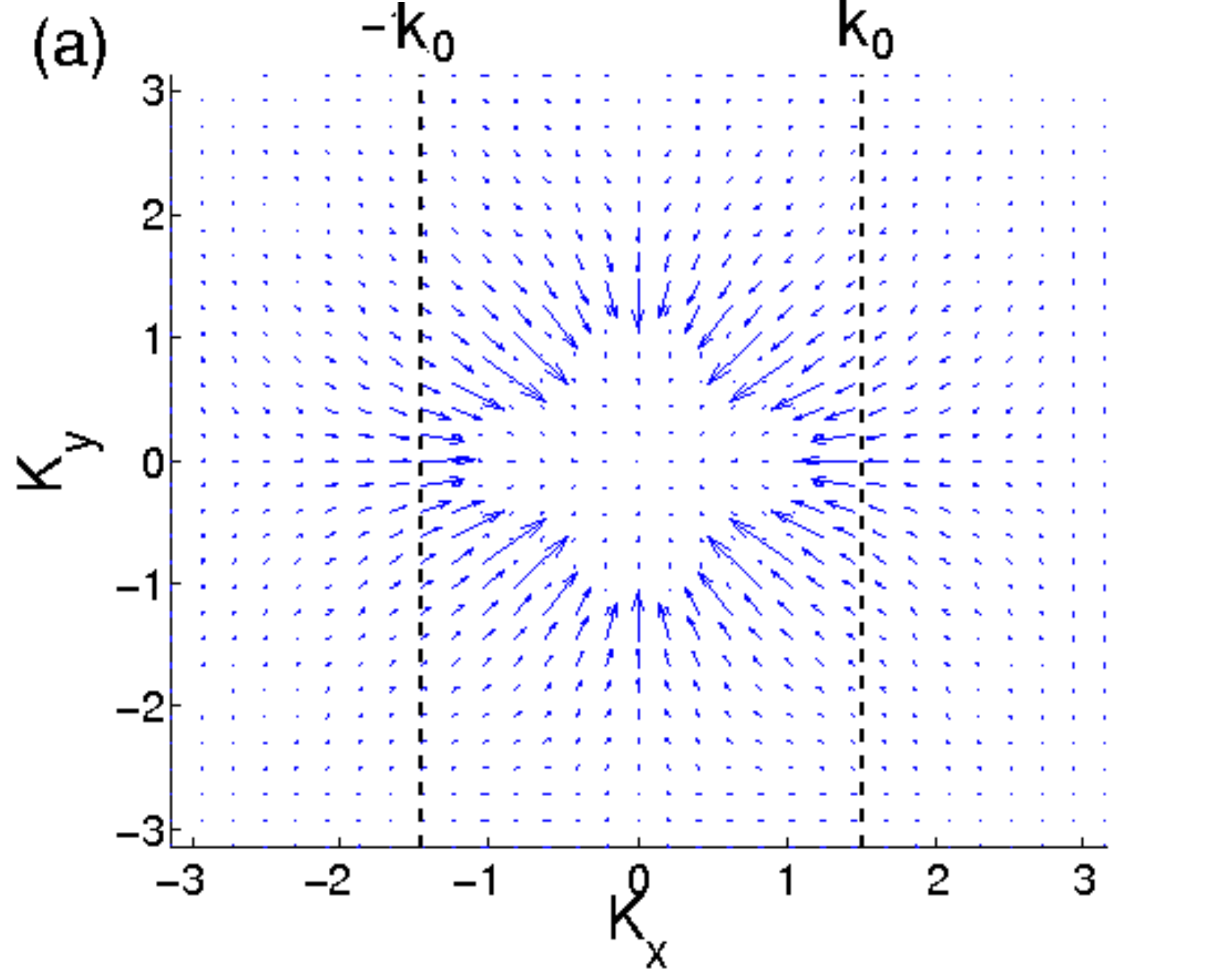}\includegraphics[width=0.41\columnwidth]{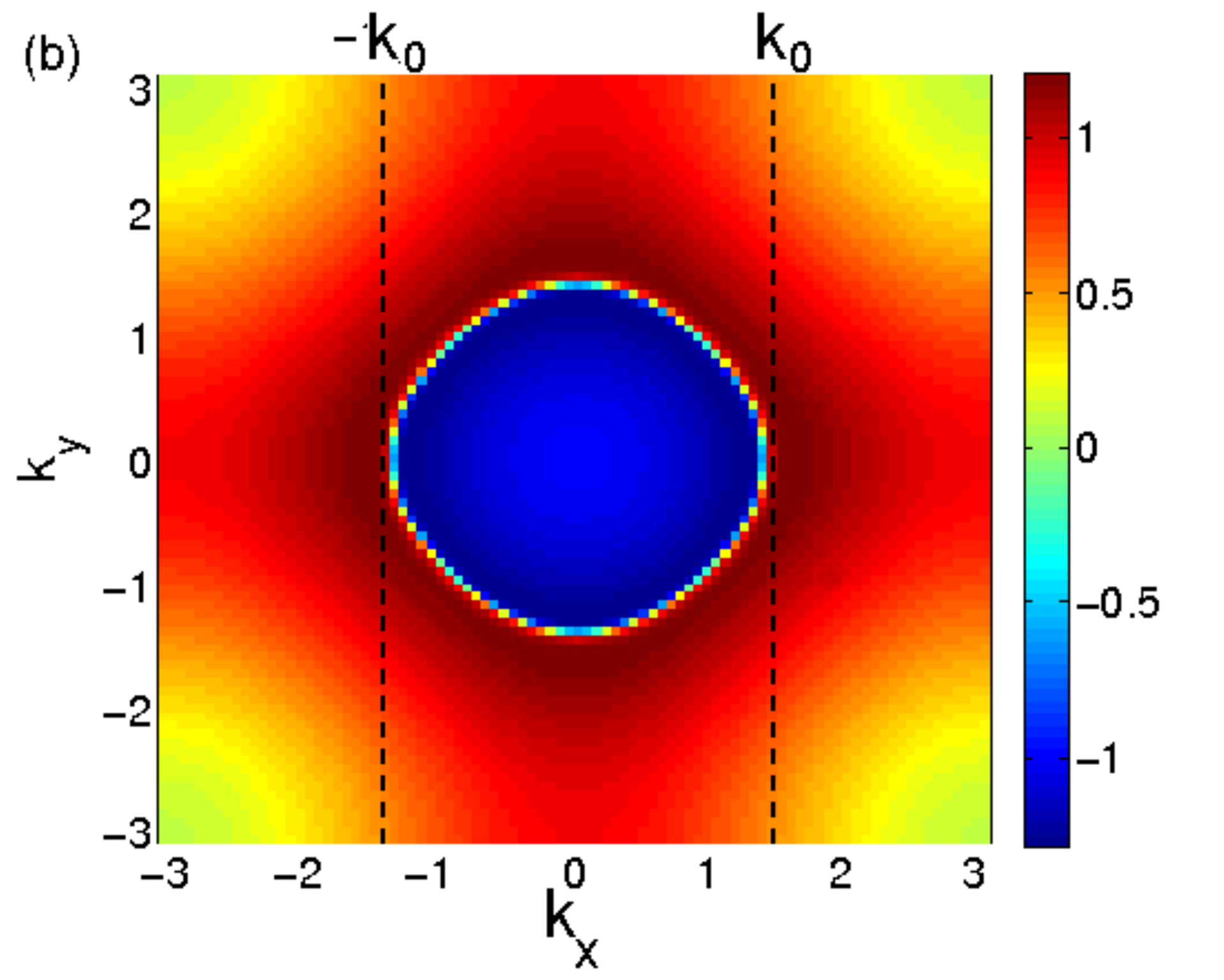}

\caption{$\vec{n}_{k}$, defined in Eq. (\ref{eq:Hf Definition}), for the
lattice model, Eqs. (\ref{eq:HgTe d(k)}) and (\ref{eq:HgTe V}),
with $\alpha=0$. \textbf{(a)} $x$ and $y$ components of $\vec{n}_{k}$
. \textbf{(b)} $n_{k}^{z}$. Note that $\vec{n}_{k}$ is in a hedgehog
configuration, as it wraps the unit sphere exactly once. This corresponds
to $C_{F}=1$. The dashed lines depict the range $[-k_{0},k_{0}]$
over which $C_{k_{x}}^{'}\neq0$.\label{Flo:n(k)}}
\end{figure}

We begin by building a description of FTIs in a generic zincblende
lattice model. The unperturbed system is given by the Bloch Hamiltonian
\begin{equation}
\begin{array}{c}
H_{k}=\left(\begin{array}{cc}
\tilde{H}_{k} & 0\\
0 & \tilde{H}_{-k}^{*}
\end{array}\right).\end{array}\label{eq:H 4x4}
\end{equation}
 This can describe, for example, $HgTe$ quantum wells, in which case
$\tilde{H}_{k}$ $\left[\left(\tilde{H}_{-k}^{*}\right)\right]$ is
a $2\times2$ Hamiltonian acting on the subspace spanned by the $J_{z}=\left(\frac{1}{2},\frac{3}{2}\right)$
$\left[J_{z}=\left(-\frac{1}{2},-\frac{3}{2}\right)\right]$ states,
respectively. Thus, the two blocks in Eq$.$ (\ref{eq:H 4x4}) are
related to each other by time reversal symmetry. Most generally, one
can write 
\begin{equation}
\tilde{H}_{k}=\vec{d}_{k}\cdot\vec{\sigma}+\varepsilon_{k}I_{2\times2}.\label{eq:H2x2}
\end{equation}
 We consider time-dependent perturbations that do not connect the
two Hamiltonian blocks, and perform the analysis on a single block.
For example, we will study the $2\times2$ Hamiltonian 
\begin{equation}
\begin{array}{c}
\tilde{H}_{lin}(t)=\vec{d}_{k}\cdot\vec{\sigma}+\varepsilon_{k}I_{2\times2}+\vec{V}_{k}\cdot\vec{\sigma}\cos\left(\omega t+\alpha\right).\end{array}\label{eq:Perturbed Hamiltonian}
\end{equation}
 where $\vec{V}_{k}$ can depict the effect of shining linearly polarized
light on the sample \cite{LRG}.

The solutions of the Schrödinger equation for a time-dependent system
evolve according to $\psi(t')=U(t',t)\psi(t)$, where $U$ is the
time evolution operator 
\begin{equation}
U\left(t',t\right)=T\left\{ \exp\left(-i\int_{t}^{t'}H\left(t''\right)dt''\right)\right\} .\label{eq:Udef}
\end{equation}
 For a time-periodic system, Floquet's theorem states that these solutions
can be written as $\psi\left(t\right)=\sum_{a}e^{i\varepsilon_{a}t}\varphi_{a}\left(t\right),$
where $\varphi_{a}\left(t\right)=\varphi_{a}(t+\tau)$ and $\tau=\frac{2\pi}{\omega}$
\cite{Eastham}. The $\varepsilon_{a}$ are called quasi-energies,
and are only defined modulo $\omega$; the $\varphi_{a}(t)$ satisfy
the eigenvalue problem $H_{F}\varphi_{a}(t)=\varepsilon_{a}\varphi_{a}(t),$
where $H_{F}$ is the Floquet Hamiltonian, obtained from the time
evolution operator over a full cycle 
\begin{equation}
e^{-iH_{F}\tau}\equiv U(\tau,0).\label{eq:e(iHft)=00003D00003DU}
\end{equation}

For a $2\times2$ block, $H_{F}$ can be written most generally as
\begin{equation}
H_{F}=\vec{n}_{k}\cdot\vec{\sigma}+\varepsilon_{k}I_{2\times2}.\label{eq:Hf Definition}
\end{equation}
 This has the structure of a gapped system provided that $\vec{n}_{k}$
does not vanish on the Brillouin zone. We then introduce a topological
invariant \cite{Bernevig,Hassan_Kane} 
\begin{equation}
C_{F}=\frac{1}{4\pi}\iint_{BZ}d^{2}k\left(\partial_{k_{x}}\hat{n}_{k}\times\partial_{k_{y}}\hat{n}_{k}\right)\cdot\hat{n}_{k}\label{eq:Skyrmion Number}
\end{equation}
where $\hat{n}_{k}=\vec{n}_{k}/\left|\vec{n}_{k}\right|$. $C_{F}$
can be nontrivial even when the unperturbed system is topologically
trivial \cite{LRG}. As a consequence, the time-dependent perturbation
can give rise to topologically-protected edge states in $H_{F}$.

The Floquet spectrum is independent of the value of $\alpha\in[0,2\pi]$.
To see this, note that a shift in $\alpha$ results in $U_{\alpha}(\tau,0)=W^{\dagger}U_{\alpha=0}(\tau,0)W$,
where $W=U_{\alpha=0}^{\dagger}(\frac{\alpha}{\omega},0)$. Thus,
changing $\alpha$ is equivalent to a similarity transformation of
$H_{F}$, $H_{F}(\alpha)=W^{\dagger}H_{F}(0)W$. The spectrum of $H_{F}$
is therefore independent of $\alpha$, even if its eigenstates are
not. In particular, varying $\alpha$ cannot induce a gap closure,
nor change $C_{F}$. Naively, this implies that at the interface between
two portions of the sample with different values of $\alpha$, no
edge modes are expected in the Floquet spectrum. However, we'll show
below configurations in which localized modes appear at such an interface.
At first glance this is remarkable, since the interface connects two
systems with identical topological classification.

\textit{Lattice model.}-- We first demonstrate the existence of quasi-stationary
interface modes in a lattice model. Consider the time-independent
Hamiltonian, Eq$.$ (\ref{eq:H2x2}), with 
\begin{equation}
\begin{array}{c}
\vec{d}_{k}=\left(A\sin k_{x},A\sin k_{y},M+2B\left(\cos k_{x}+\cos k_{y}-2\right)\right).\end{array}\label{eq:HgTe d(k)}
\end{equation}
We choose $\varepsilon_{k}=0$. Then Eq$.$ (\ref{eq:H2x2}) has particle-hole
symmetry (PHS) in which the valence and conduction bands are interchanged.
We add a time-dependent perturbation 
\begin{equation}
V=V_{0}\sigma_{z}\cos\left(\omega t+\alpha\right).\label{eq:HgTe V}
\end{equation}

\begin{figure}[t]
\includegraphics[width=0.375\columnwidth]{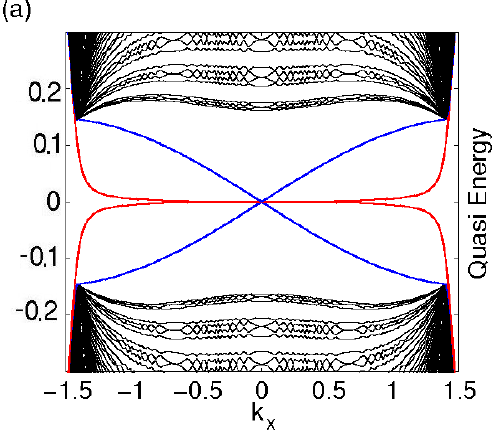}\includegraphics[width=0.44\columnwidth]{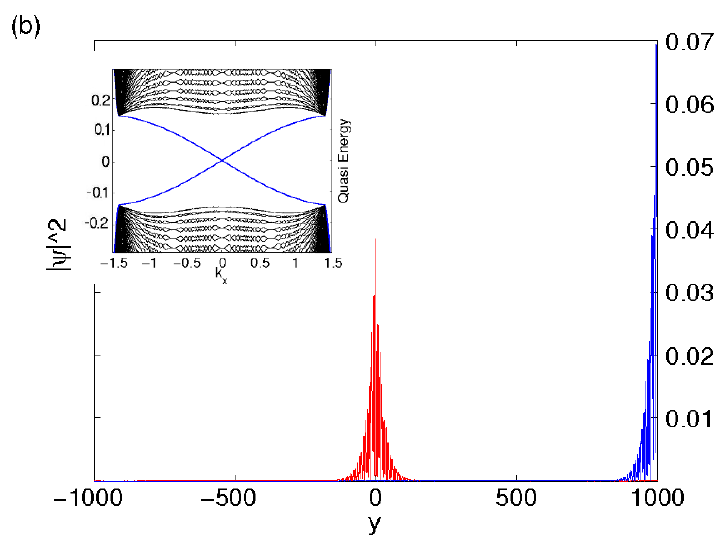}

\caption{\textbf{(a) }Floquet spectrum of Eq$.$ (\ref{eq:Perturbed Hamiltonian})
for a domain wall configuration. The plotted range\textbf{ }corresponds
approximately to\textbf{ $\left[-k_{0},k_{0}\right]$.(b)} Amplitude
of the localized bulk (red) and edge (blue) states \label{Flo:Numerical Spectrum Bulk}.
\textbf{Inset}: Floquet spectrum for constant $\alpha$. Results are
for $V_{0}=A=-B=0.2M,\omega=2.7M$. }
\end{figure}

\begin{figure}[t]
 \includegraphics[width=0.7\columnwidth]{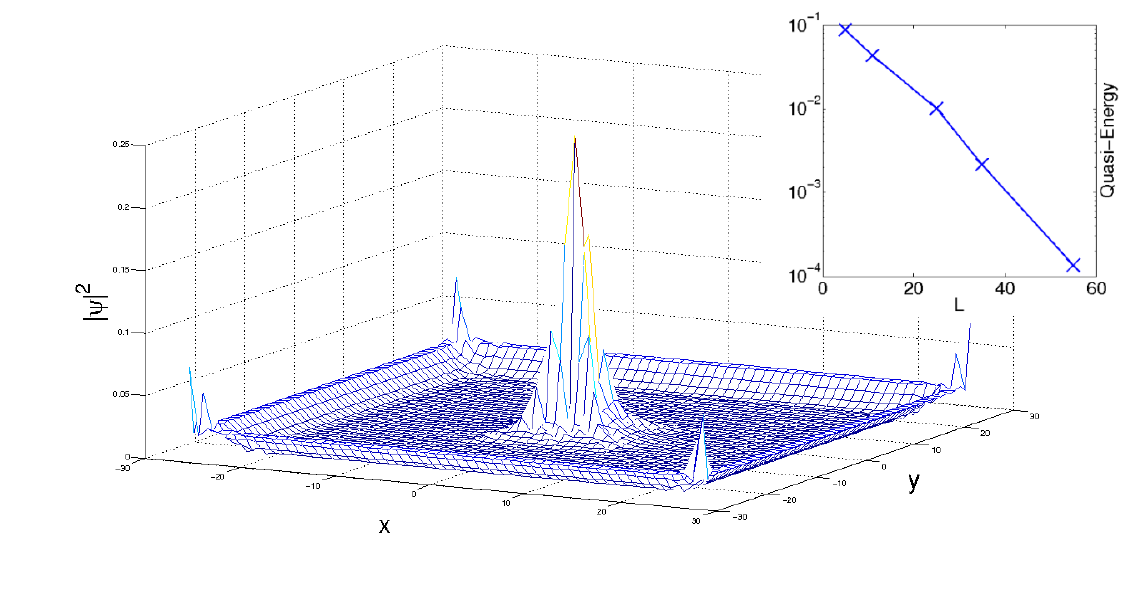}\caption{Amplitude of the vortex bound state. \textbf{Inset:} quasi-energy
of the state as a function of the system size $L$. \label{Flo:Vortex Numerical Spectrum}
Results are for $V_{0}=1.5M,A=0.2M,B=-0.1M,\omega=2.7M$}
\end{figure}

We evaluate the time evolution operator numerically, Eq$.$ (\ref{eq:Udef}),
by discretizing the time interval between $t=0$ and $t'=\tau\equiv\frac{2\pi}{\omega}$.
We then obtain the Floquet spectrum by diagonalizing $U$, see Eq$.$
(\ref{eq:e(iHft)=00003D00003DU}). When $\alpha$ is a constant, the
system has translational invariance and $\vec{n}_{k}$ can be calculated
for each $\vec{k}$ separately. Fig. \ref{Flo:n(k)} shows that $\hat{n}_{k}$
wraps exactly one time around the unit sphere over the first Brillouin
zone. The system is therefore topologically non-trivial with $C_{F}=1$,
and we expect one localized chiral mode at each edge of the system.
To demonstrate this, we choose cylindrical geometry with open boundaries
in the $y$ direction and periodic boundaries in the $x$ direction.
Figure \ref{Flo:Numerical Spectrum Bulk} shows that indeed, one zero
quasi-energy mode exists at each boundary of the system. This corresponds
to the edge state found in Ref. \cite{LRG}.

We now allow $\alpha$ to be position-dependent. As a first example,
we consider a domain wall configuration, across which the external
perturbation changes sign, $\alpha\left(y\right)=\pi\theta\left(y\right)$.
This captures the phase shift across the nodes of a standing wave
created by two interfering light rays incident on the sample. By adjusting
the incidence angle of the rays, one can control the periodicity of
the standing wave such that these nodes are well-separated. We choose
cylindrical geometry. Since the system remains translationally invariant
along $x$, we work in the hybrid coordinate basis $\left(k_{x},y\right)$
and diagonalize $U$ for each $k_{x}$ value. Figure \ref{Flo:Numerical Spectrum Bulk}
shows the resulting quasi-energy spectrum. Note that, in addition
to the edge modes, the spectrum now includes two counter-propagating
zero quasi-energy modes localized near the domain wall at $y=0$.

As a second example, we consider a vortex configuration, in which
the phase $\alpha$ winds by $2\pi$ about a point, $\alpha\left(\vec{r}\right)=\arctan(y/x)$.
A lattice of such vortices can be created by interfering three lasers,
and an isolated vortex can be created using a phase mask. We set open
boundaries and diagonalize $U$ in real space to obtain the Floquet
spectrum. In addition to the edge modes, we find a zero quasi-energy
state bound to the vortex. Figure \ref{Flo:Vortex Numerical Spectrum}
shows the wave function of the bound state. This state hybridizes
with an edge state and opens a small gap. However, this is a finite-size
effect, and the gap energy decays exponentially with system size.

\textit{Analogy to $p_{x}+ip_{y}$ superconductors.--} In order to
explain these results, we establish an analogy between our system
and a $p_{x}+ip_{y}$ superconductor (pSC). We then relate the domain
wall and vortex core states to the well-known zero energy states of
pSCs.

We derive an approximate analytic expression for  $H_{F}$. The derivation
is first carried out for constant $\alpha$. We start by omitting
the component of $\vec{V}_{k}$ parallel to $\hat{d}_{k}$, since
it only affects the dynamics weakly when averaged over a full cycle
and is known not to influence the topological properties of the system
\cite{LRG}. The remaining perpendicular component is $\vec{V}_{k,\perp}=\vec{V}_{k}-\left(\vec{V}_{k}\cdot\hat{d_{k}}\right)\hat{d_{k}}$.
We define $V_{k,\perp}\equiv\left|\vec{V}_{k,\perp}\right|$ and introduce
$\hat{v}_{k}=\vec{V}_{k,\perp}/V_{k,\perp}$ and $\hat{w}_{k}=\hat{d}_{k}\times\hat{v}_{k}$,
such that $\hat{d}_{k},\hat{v}_{k}$ and $\hat{w}_{k}$ form a right-handed
triad. The perturbation can be decomposed into terms that rotate and
counter-rotate about $\hat{d}_{k}$, 
\[
\begin{array}{c}
\tilde{H}_{k}\approx\vec{d}_{k}\cdot\vec{\sigma}+\frac{1}{2}V_{k,\perp}\left(\hat{v}_{k}\cos\omega t+\hat{w}_{k}\sin\omega t\right)\cdot\vec{\sigma}\\
+\frac{1}{2}V_{k,\perp}\left(\hat{v}_{k}\cos\omega t-\hat{w}_{k}\sin\omega t\right)\cdot\vec{\sigma}.
\end{array}
\]
 We go to a rotating frame through the time-dependent unitary transformation
$R\left(t\right)=\exp\left(-i\hat{d}_{k}\cdot\vec{\sigma}\frac{\omega t}{2}\right)$.
The resulting states $\left|\psi(t)\right\rangle _{r}=R(t)\left|\psi(t)\right\rangle $
satisfy $i\partial_{t}\left|\psi(t)\right\rangle _{r}=\tilde{H}'\left|\psi(t)\right\rangle _{r},$
where $\tilde{H}'=R^{\dagger}\left(\tilde{H}-iI\partial_{t}\right)R$.
We then find
\[
\begin{array}{c}
\tilde{H}'_{k}=\left(\vec{d}_{k}-\frac{\omega}{2}\hat{d}_{k}+\frac{1}{2}V_{k,\perp}\left(\hat{v}_{k}\cos\alpha+\hat{w}_{k}\sin\alpha\right)\right)\cdot\vec{\sigma}\\
+\frac{1}{2}V_{k,\perp}\left(\hat{v}_{k}\cos\left(2\omega t+\alpha\right)-\hat{w}_{k}\sin\left(2\omega t+\alpha\right)\right)\cdot\vec{\sigma}.
\end{array}
\]
 The Rotating Wave Approximation (RWA) consists of neglecting the
$2\omega$ term in $\tilde{H}_{k}^{\prime}$. This is valid near a
resonance, $\left|\omega-\Delta E\right|\ll\omega+\Delta E$, where
$\Delta E$ is the energy difference between states in the lower and
upper bands. This procedure yields a time-independent operator 
\begin{equation}
\begin{array}{c}
H_{F,\alpha}^{RWA}=\left(\left(1-\frac{\omega}{2d_{k}}\right)\vec{d}_{k}+\frac{1}{2}V_{k,\perp}\left(\hat{v}_{k}\cos\alpha+\hat{w}_{k}\sin\alpha\right)\right)\cdot\vec{\sigma},\end{array}\label{eq:TI Floquet Hamiltonian}
\end{equation}
 which is the Floquet Hamiltonian in the RWA.

The analogy to a pSC can be seen explicitly by performing a $k$-dependent
unitary transformation that rotates $(\hat{d}_{k},\hat{v}_{k},\hat{w}_{k})\to(\hat{z},\hat{k},\hat{z}\times\hat{k})$,
where $\vec{k}=(k_{x},k_{y})$. This leads to a Hamiltonian of the
form

\begin{equation}
H_{F}^{'}=\left(\begin{array}{cc}
\zeta_{k} & \Delta_{k}e^{-i\alpha}\left(k_{x}-ik_{y}\right)\\
\Delta_{k}e^{i\alpha}\left(k_{x}+ik_{y}\right) & -\zeta_{k}
\end{array}\right),\label{eq:P-SC Analogue Hf}
\end{equation}
where $\zeta_{k}=d_{k}-\omega/2$ and $\Delta_{k}=V_{k,\perp}/2k$
are real. Equation (\ref{eq:P-SC Analogue Hf}) resembles the Hamiltonian
of a pSC with complex order parameter $\Delta_{k}e^{-i\alpha}$. The
analogy to the pSC can be seen graphically in Fig$.$ \ref{Flo:n(k)},
where panel (a) depicts the normal dispersion and panel (b) the superconducting
order parameter, which is seen to have $p_{x}+ip_{y}$ symmetry. Note
that, unlike an actual superconductor in which the Nambu basis describes
particle and hole states, here Eq$.$ (\ref{eq:P-SC Analogue Hf})
acts on two particle-like states corresponding to valence and conduction
bands of the Floquet problem. Hence, the spectrum of Eq$.$ (\ref{eq:P-SC Analogue Hf})
matches the corresponding pSC, but the nature of the wave functions
in the two cases is related by a particle-hole transformation.

We apply these results to our system. The unperturbed Hamiltonian
for small $\vec{k}$ is $\widetilde{H}_{k}=A\vec{k}\cdot\vec{\sigma}+M\sigma_{z}.$
The Floquet Hamiltonian in the RWA is then
\[
\begin{array}{c}
H_{F,\alpha}^{RWA}=\left(\eta\left(\hat{z}+\frac{A}{M}\vec{k}\right)+\Delta_{0}\left(\vec{k}\cos\alpha+\hat{z}\times\vec{k}\sin\alpha\right)\right)\cdot\vec{\sigma}\end{array}
\]
where $\eta=\left(M-\frac{\omega}{2}\right)$ and $\Delta_{0}=\frac{AV_{0}}{2M}$.
When $\alpha$ varies in space, this is generalized to a Bogoliubov-de
Gennes (BdG) equation (here $\psi=(u,v)^{T}$): 
\begin{equation}
\begin{array}{c}
\left(\varepsilon-\eta\right)u=\left(\frac{A}{M}\eta-\Delta\left(\vec{r}\right)\right)*\left(-i\partial_{x}+\partial_{y}\right)v\\
\left(\varepsilon+\eta\right)v=\left(-i\partial_{x}-\partial_{y}\right)*\left(\frac{A}{M}\eta-\Delta^{*}\left(\vec{r}\right)\right)u
\end{array}\label{eq:BdG}
\end{equation}
 where $*$ denotes the symmetric product $a*b=\frac{1}{2}\left(ab+ba\right)$.
In theories of superconductivity the gap function $\Delta\left(\vec{r}\right)$
is calculated self-consistently. Here, $\Delta\left(\vec{r}\right)$
is directly determined by the external perturbation.

When $\alpha\left(\vec{r}\right)$ describes a vortex, the BdG equation
has a zero energy solution bound to the vortex core. This state is
analogous to the well-known pSC vortex core modes, and just as in
the case of a pSC, it is topologically protected provided particle-hole
symmetry is present \cite{Read &Green,TeoKane}. This gives a fractional
charge of $\pm1/2$ at the vortex, depending on whether the state
is occupied or not \cite{Chamon}. In the full $4\times4$ system,
the vortex is felt in both subspaces, yielding two zero energy states
which can be independently occupied. This leads to fractionalized
excitations $(\pm1/2,\pm1/2)$, where the indices denote the charge
in each subspace. The scenario is closely analogous to fractionalization
in polyacetylene, where fractional charge excitations in the spin-up
and spin-down channels combine into spin $1/2$ excitations with integer
charge \cite{Su Schrieffer Heeger}. 

When $\alpha\left(\vec{r}\right)$ describes a domain wall, the BdG
equation has two zero quasi-energy solutions, localized around $y=0$
with a localization length of $\xi=\frac{\eta}{\Delta_{0}}$. This
configuration is analogous to an interface between two pSCs with a
relative phase difference of $\pi$ -- a {}``$\pi$ junction'' --
which is known to have zero energy bound Andreev states \cite{Kitaev,TeoKane}.
To understand these modes, let's first consider a system with no disorder
and no interactions. Then $k_{x}$ is a good quantum number and the
system reduces to a 1D pSC for each $k_{x}$. We can then define a
$k_{x}$-dependent topological invariant, $C_{k_{x}}^{'}$, as the
winding number of $\hat{n}_{k}$ in the $(n_{k}^{y},n_{k}^{z})$ plane.
As can be seen from Fig. \ref{Flo:n(k)}, when $\alpha=0$ the 1D
pSC is topological with $C_{k_{x}}^{'}=1$ for $k_{x}\in\left[-k_{0},k_{0}\right]$
and it is trivial otherwise. On the other side of the domain wall,
$\alpha=\pi$, and the sign of $C_{k_{x}}^{'}$ is reversed. For $k_{x}\in\left[-k_{0},k_{0}\right]$
the change in sign in $C_{k_{x}}^{'}$ implies a pair of localized
states exists at the interface, which disperse in opposite directions
as a function of $k_{x}$. In particular, at $k_{x}=0$ these modes
cross with zero quasi-energy. When PHS is present, one of these states
is even under the PH transformation, whereas the other is odd, preventing
them from mixing and opening up a gap. This protection is robust even
in the presence of disorder and interactions, provided these do not
break PHS.

Note that the domain wall and vortex configurations are closely related.
To see this, we write $\alpha\left(\vec{r}\right)=\arctan\left(y/\beta x\right)$,
which describes a vortex for $\beta=1$ and a domain wall for $\beta\to0^{+}$.
Thus, the domain wall is a continuous deformation of a vortex and
the $\pi$ jump arises from the vortex winding. The interface modes
are therefore smoothly connected to the vortex core states.

\textit{Photoinduced current.}-- Recall that imposing a phase twist
on a superconductor $\Delta e^{i\alpha\left(\vec{r}\right)}$ results
in a Josephson current $\vec{j_{S}}=\rho_{s}\vec{\nabla}\alpha$ \cite{Schrieffer}.
Motivated by the analogy to pSC, we consider the effect of a slowly-varying
phase twist $\alpha\left(y\right)=\alpha_{0}+y\partial_{y}\alpha$,
where $\partial_{y}\alpha$ is a small constant. This can be achieved,
for example, by shining a coherent light ray incident at an angle
to the surface of the sample. Indeed, we find that the system experiences
a DC current along $\hat{y}$, and compute an analogue of the superfluid
stiffness $\rho_{s}$ in terms of the vectors $\vec{d}_{k}$ and $\vec{n}_{k}.$

By Noether's theorem, the current density operator is 
\begin{equation}
\vec{j}=-\vec{\nabla}_{k}\left(\vec{d}_{k}\cdot\overrightarrow{\sigma}\right)\label{eq:current operator}
\end{equation}
 where $\vec{\nabla}_{k}=\left(\partial_{k_{x}},\partial_{k_{y}}\right)$.
Our goal is to compute the expectation value of Eq$.$ (\ref{eq:current operator})
with respect to the state at half-filling, in which all eigenstates
in the Floquet valence band are fully occupied,
\begin{equation}
\left\langle \vec{j}\right\rangle =\sum_{\psi_{k}}\raisebox{-1.4mm}{\mbox{\ensuremath{{\scriptstyle r}}}}\left\langle \psi_{k}\right|R\left(t\right)\hat{j}R^{\dagger}\left(t\right)\left|\psi_{k}\right\rangle _{r}\label{eq:current EV}
\end{equation}
 where the sum is over all the negative quasi-energy states. Here,
we insert $R\left(t\right)=\exp\left(-i\hat{d}_{k}\cdot\vec{\sigma}\frac{\omega t}{2}\right)$
since the current is computed in the lab frame.

In order to derive an expression for the current, we compute $H_{F}$
to linear order in $\partial_{y}\alpha$ in the RWA. We obtain $H_{F}=H_{F,\alpha_{0}}^{RWA}+(\partial_{y}\alpha)H_{1}^{RWA}$,
where $H_{F,\alpha_{0}}^{RWA}$ is given by Eq$.$ (\ref{eq:TI Floquet Hamiltonian})
and 
\[
H_{1}^{RWA}=\frac{y}{4}V_{k,\perp}\left(\cos\alpha_{0}\hat{w}_{k}-\sin\alpha_{0}\hat{v}_{k}\right)\cdot\overrightarrow{\sigma}+h.c.
\]
 We then write $|\psi_{k}\rangle_{r}=|\psi_{k}^{0}\rangle_{r}+\partial_{y}\alpha|\psi_{k}^{1}\rangle_{r}$,
where $|\psi_{k}^{0}\rangle_{r}$ are eigenstates of $H_{F,\alpha_{0}}^{RWA}$,
and $|\psi_{k}^{1}\rangle_{r}$ is obtained from $1^{st}$ order perturbation
theory. The resulting current, to $O\left(\partial_{y}\alpha\right)$,
vanishes along $\hat{x}$, while along $\hat{y}$ it is:
\[
\begin{array}{c}
\frac{\langle j_{y}\rangle}{\partial_{y}\alpha}=\iint\frac{d^{2}k}{(2\pi)^{2}}\frac{V_{0}}{n_{k}}\left(\hat{d}_{k}\times\nabla_{k}\vec{d}_{k}\right)\cdot\hat{v}_{k}\,\raisebox{-1.4mm}{\mbox{\ensuremath{{\scriptstyle r}}}}\left\langle \psi_{k}^{0}\right|i\overleftrightarrow{\partial}\left|\psi_{k}^{0}\right\rangle _{r}\end{array}
\]
where $\left\langle \psi\right|\overleftrightarrow{\partial}\left|\psi\right\rangle =\left\langle \psi\right|\left.\partial\psi\right\rangle -\left\langle \partial\psi\right|\left.\psi\right\rangle $.
In this expression, we have summed over the contributions coming from
both sub-Hamiltonians, $\tilde{H}_{k}$ and $\tilde{H}_{-k}^{*}$,
appearing in Eq$.$ (\ref{eq:H 4x4}). In terms of $\vec{d}_{k}$
and $\vec{n}_{k}$, this can be rewritten as 
\begin{equation}
\begin{array}{c}
\frac{\langle j_{y}\rangle}{\partial_{y}\alpha}=\iint\frac{d^{2}k}{(2\pi)^{2}}\frac{V_{0}}{2}\frac{d_{k}}{n_{k}}\frac{\hat{n}_{z}}{\hat{n}_{x}^{2}+\hat{n}_{y}^{2}}\left(\left(\hat{d}_{k}\times\partial_{k_{y}}\hat{n}_{k}\right)\cdot\hat{z}\right)^{2}\end{array}\label{eq:lab frame Current}
\end{equation}
 where $d_{k}=\left|\vec{d}_{k}\right|$ and $n_{k}=\left|\vec{n}_{k}\right|$.

Integrating Eq$.$ (\ref{eq:lab frame Current}) over the Brillouin
zone gives a non-zero DC current. As a check of our analysis, we compared
these results with numerical simulations on the lattice model and
found good agreement. This result is reminiscent of the photogalvanic
effect proposed to exist at the surface of 3D topological insulators
radiated with circularly polarized light \cite{Hosur}.

\textit{Discussion.--} The analogy to pSC relies on particle-hole
symmetry. Thus, it is natural to ask which of our results rely on
this symmetry. For instance, Eq$.$ (\ref{eq:lab frame Current})
for the photoinduced current is valid even when PHS is broken. In
contrast, PHS plays an important role in preventing the interface
modes from opening a gap, as discussed above. For real systems PHS
is only approximate, and we must consider the effects of weak PHS
breaking. We find that this induces a small mixing of the interface
modes. For example, using the same parameters as in Fig. \ref{Flo:Numerical Spectrum Bulk},
but adding a PHS breaking term to Eq$.$ (\ref{eq:Perturbed Hamiltonian})
of the form $\varepsilon_{k}=-0.2\left(\cos k_{x}+\cos k_{y}-2\right)$,
we find that a small gap $\approx10^{-5}$ opens up. Thus, experiments
carried out at temperatures above this gap will not be sensitive to
PHS breaking. The vortex core state shows higher degree of robustness
to breaking of PH symmetry. Although PH symmetry breaking shifts the
bound state quasi-energy away from zero, it remains a mid-gap state
provided the symmetry is weakly broken. In particular, the fractional
nature of the excitation is unaffected.

We now briefly discuss the physical manifestations of our results
in realistic systems. Earlier work has shown that resonantly-driven
systems can reach a steady state, with occupation given by the Fermi-Dirac
distribution based on the Floquet spectrum, and with an effective
temperature that is determined by interaction and phonon relaxation
rates \cite{Glazman,Eliashberg}. Then, a standing wave pattern is
expected to create one dimensional channels along the domain walls
and thus induce anisotropic conductivity. It may be possible to generalize
our results to systems in which the light frequency exceeds the bandwidth
(that is, systems that are driven off-resonance). The advantage of
doing this is that physical properties of systems driven off-resonance
are much easier to understand \cite{Fertig Gu Arovas Auerbach,Oka_Aoki,Kitagawa_Oka_Brataas_Fu_Demler}.
Modulated FTIs driven off-resonance, and in three dimensions, will
be discussed in future work.

In conclusion, we provided a set of schemes by which the properties
of Floquet topological insulators can be manipulated using modulated
light. We proposed explicit setups by which bulk $1D$ channels, fractionalized
excitations, and light induced currents can be generated. Our analysis
demonstrates the great potential for tunability and control of light-induced
topological phases.

\textit{Acknowledgments.-- }We would like to thank A$.$ Auerbach,
V. R. Chandra, E$.$ Demler, L. Glazman, N$.$ Lindner, A$.$ Paramekanti,
and G$.$ Refael for many useful discussions. Part of this work was
done during a visit to the Aspen Center for Physics. This research
was supported by the Israeli Science Foundation under grant 1338/09
and by a Marie Curie IRG grant.

\end{document}